# Simulation Study for the Energy Resolution Performances of Homogenous Calorimeters with Scintillator-Photodetector Combinations

G. Aydın[1]

[1] Department of Physics, Mustafa Kemal University, TR-31034, Hatay, Turkey

Correspondence should be addressed to G. Aydın; gaydin@mku.edu.tr

**Abstract**

The scintillating properties of active materials used in high energy and particle physics experiments play an important role regarding the performances of both calorimeters and experiments. Two scintillator materials, a scintillating glass and an inorganic crystals were examined to be used for collider experiments showing good optical and scintillating properties. This paper discusses the simulated performances of two materials of interest assembled in a scintillator-photodetector combination. The computational study was carried out with Geant4 simulation program to determine energy resolutions of such calorimeter with different beam energies and calorimeter sizes.

## Introduction

Scintillator materials are used in high energy physics experiments as active materials of calorimeters to measure energy and position of particles passing through calorimeters generating photons proportional to incoming beam energies. Two type of calorimeters could be constructed: sampling or homogenous [1, 2]. A sampling calorimeter consists of an absorber and an active material in alternating layers resulted in absorption of only some part of incident beam energy in active materials. A homogenous calorimeter is entirely made of an active material with no absorber, thus leading to absorption of the most of incident energy depending on thickness and radiation length of the material. Sampling calorimeters serve for both electromagnetic and hadronic interactions but homogenous calorimeters are used for only electromagnetic interactions due to their long interaction length. In such detector systems, several properties of active materials affect the performance of calorimeters and experiments. First of all, light yield of a scintillator should be high enough to obtain required energy resolution. Next, the rate of data taking is important when considering short time intervals between collisions. Therefore, the response time of detectors should be as fast as possible to detect even rare events. Decay times of scintillator materials affect the time interval of signal formation and thus they are key factors for data taking rate in calorimeters. High density in scintillating materials increase stopping power and it is important in two ways. One is that it increases energy and spatial resolutions and it enables construction of more compact systems. Moreover, a scintillator with a good optical transmission has significant impact on the formation of proper electrical signal in photo-detectors to which photon pulses produced in the scintillator are directed. Beyond these facts, scintillator materials could show very good properties in some features but could also have some drawbacks for the remaining aspects. In summary, scintillating materials with high densities, required light yield, producing fast and short light pulses have crucial role to build detector systems which are compact, enabling fast data taking, and achieving required energy and spatial resolutions. For example, lead tungsten crystals (PWO) with high density and fast decay times are used in electromagnetic part of the Compact Muon Selenoid (CMS) to measure incident electron or photon beam energies which is used to search Higgs boson [1, 2]. New experiments are also searching for scintillating materials with good optical and scintillating properties as much as possible as an active material in calorimeter designs. This report presents a computational study concentrated on energy resolution performances of different scintillating materials which could be used as an active material of a homogenous calorimeter in particle physics experiments. The interested materials are Ce doped HfG (Hafnium Fluoride Glass) [3,4] and Ce doped $Gd_2Y_1Ga_{2.7}Al_{2.3}O_{12}$ [5] due to their good optical and scintillating properties. Here, HfG is a scintillating glass and the other is inorganic crystal. To the best to our knowledge, both scintillating materials were not used in a high energy physics experiment as an active material of a homogenous calorimeter or their simulation studies were not presented for energy resolution calculation belonging to a certain size or sets of calorimeter setups. These materials have mass production capabilities. Generally, scintillation glasses are potentially more homogenous compared to crystal scintillators and light yield of scintillation glasses could be increased by changing their elemental compositions. The selected glass material has been preferred among some heavy metal fluoride glass due to its optical and scintillation properties. On the other hand, these improvements in crystals could be achieved by increasing purity of the crystals and with better understanding of luminescence mechanisms. Scintillation glasses have less light yield compared to crystals but generally fast decay times [6].

This study determines energy resolution of homogenous calorimeter setups with certain sizes to see the performance of the selected materials. The detailed explanation of the materials belonging to their physical and scintillation properties and the simulation procedure are given in detail in Section 2.

## Materials and Methods

The energy resolution of a scintillator could be characterized with four parameters: Lateral part, photostatistics contribution, constant, and noise term. The energy resolution is the quadratic summation of all four terms as indicated in Equation 1.

$$\frac{\sigma(E)}{E} = \frac{a_{lateral}}{E^{1/4}} \oplus \frac{a_{pe}}{\sqrt{E}} \oplus b \oplus \frac{c}{E} \qquad (1)$$

Here, $a_{lateral}, a_{pe}, b,$ and $c$ refer to lateral part, photoelectron statistics contribution, constant, and noise terms, respectively. The lateral part represents fluctuations of shower development inside the scintillating material ($a_{lateral}$) belonging to lateral shower containment and contribution from the statistics of the photoelectrons produced in a photodetector, which converts photons reaching its active area into electrons in terms of its wavelength dependent quantum efficiency and its internal gain, is represented with $a_{pe}$. The total energy resolution is calculated as the quadratic summation of all terms excluding noise term in this study. The constant term refers to other inhomogeneities of the material such as variations of reflectance of different surfaces of the scintillator or impurities in the material. This is also some portion of the energy deposition fluctuation. The contribution from photoelectron statistics is given by Equation 2 [2].

$$a_{pe} = \sqrt{\frac{\bar{F}}{N_{pe}}} \qquad (2)$$

Here, $N_{pe}$ is the the number of photoelectron per GeV and F is the emission weighted excess noise factor due to avalanche gain process. $N_{pe}$ depends on the number of photons reaching the rear edge of the scintillator, active area of photodetectors, and wavelength dependent quantum efficiencies of the photodetectors. In the presented study, noise term is not included in energy resolution calculation. Four different photodetectors were used in the study: two PIN diodes and two avalanche photo diodes (APD). Recently, several experiments used the pin diode: Babar [7], BELLE [8], and BesIII [9] used Hamamatsu S2744-08 PIN diode to collect photons from crystals. APD S8664-55, on the other hand, are used in CMS [10] with PWO calorimeter. The pin diode S2744-08 and APD S8664-55 have spectral ranges of 340 to 1100 nm and 320 nm and 1100 nm, respectively [11]. These were used with $Gd_2Y_1Ga_{2.7}Al_{2.3}O_{12}$ scintillator. Another Si APD and PIN diode were used for HfG by considering their emission spectra which peak at relatively lower wavelength. It is Si APD S5345 whose spectral range is between 200 nm and 1000 nm and Si PIN diode S1227-1010BQ with 190 to 1000 nm spectral range [11]. The photodetectors used in the presented study have the active areas of 1 cm x 2 cm, 5 mm x 5 mm, and 5 mm in diameter, and 10 mm x 10 mm for S2744-08 pin diode, APD S8664-55, and ADP S5345, and S1227-1010BQ pin, respectively. The quantum efficiency of the pin diode S2744-08 is around 10% at the wavelength of 300 nm, 50% at 400 nm, and reaches 83% at 580 nm wavelength of the photon emission. On the other hand, those efficiency values are seen for APD S8664-55 as 23%, 70%, and 85% indicating obviously that APD is more efficient at relatively higher wavelengths. Si photodiodes have fast response, high sensitivity and low noise. PIN diodes have no internal gain, so they do not contribute to photoelectron statistics due to excess noise factor. Since APD has avalanche gain process, it contributes to the photoelectron statistics term as excess noise factor due to fluctuations in gain process. This factor is wavelength dependent and the excess noise factor is determined as 2 for the emission wavelength below 500 nm [12]. This value was used for HfG scintillator - photodetector systems. On the other hand, this factor was calculated as 2.346 for the $Gd_2Y_1Ga_{2.7}Al_{2.3}O_{12}$ by taking account its emission spectrum together with excess noise factor distribution of an APD as a function of wavelength [12] according to the Equation 3.

$$\bar{F} = \frac{\int F(\lambda) Em(\lambda) d\lambda}{\int Em(\lambda) d\lambda} \qquad (3)$$

Where, $F(\lambda)$ is the excess noise factor as a function of wavelength and $Em(\lambda)$ is the emission weights of the spectrum for a given scintillator material.

The scintillating process in Geant4 [13-15] is following: energy lost for each step determines the number of optical photons which has Gaussian distribution shape and statistical fluctuations occur around average light yield entered as scintillation yield. Photons are generated along beam direction emitted uniformly into 4π. They are emitted according to random linear polarization and scintillation time components. This process produces optical photons which are directly used in a desired application. Instead of working directly with optical photons, the following method is reasonable by taking account required information belonging the material and electronics coupled to it.

The number of optical photons produced in a scintillation process is proportional to energy deposition in the material during the process. If there is no self-absorption in the material, these produced photons will be directed to the rear back of the material in theoretical limits. Therefore, the following procedure to calculate energy resolution is an appropriate method by using Geant4 program: First of all, the distribution of energy deposition event to event is obtained for the related beam energies. Fitting this distribution with suitable function will give energy resolution value for the interested calorimeter setup. This value refers to the contribution to energy resolution due to event to event fluctuations of the number of produced optical photons. The remaining part of the fluctuation is due to photoelectron statistics. In this study, since pin diodes have no internal gain, the contribution from variance of the gain process in the photodetector is neglected for pin diodes. Here, the calculation of the average number of photoelectrons produced in the photodetectors determines the photoelectron statistics contribution. Clearly, it depends on the number of photons produced in the material and reaching to the active areas of the photodetectors. The light yield is the main scintillation property of a scintillating material. In this study, the yield values per MeV were used to determine the average number of photons produced in the optical process. Then, the number of photons reaching the rear edge of the material was calculated by taking account of transmission spectra of the interested materials. The next step was to determine the number of photons hitting the photodetector active area by taking account of the ratio of the photodetector area to the total back face area of the scintillator. The final step was to determine the number of photoelectrons produced in the photodetectors according to their emission weighted quantum efficiencies.

In the present study, two beam facing areas (20 mm x 20 mm and 25 mm x 25 mm) were selected for each scintillator forming 5 x 5 matrix geometry. In this way, the total beam facing area was either 100 mm x 100 mm or 125 mm x 125 mm. The areas of beam facing and back face of each scintillator were set equally in the simulation. Five different thicknesses (17 cm, 20 cm, 23 cm, 25 cm, and 27 cm) were tested for calorimeter performances. Totally, ten geometric configurations were examined to see the changes of calorimeter performances with detector sizes and to compare obtained results with previous experimental or simulation results in certain sizes. Gamma was used as an incident beam with different energies ranging from 100 MeV to 2 GeV by directing the beam to the center of the matrix. The ratios of the active area of the APDs S8664-55 and S5345 to the total area of the calorimeter back face were calculated as 0,125 and 0,0982 for 20 mm x 20 mm and 0,08 and 0,0628 for 25 mm x 25 mm back face areas of the detectors, respectively taking account that each scintillators locates a pair of diodes at the back faces. These values were calculated for pin S2744 and pin S1227 as 1,0 and 0,5 with 20 mm x 20 mm and 0,64 and 0,32 with 25 mm x 25 mm back face areas of the detectors, respectively.

The previous studies with CsI(Tl), PWO, and LYSO crystals showed that the mentioned simulation procedure gives compatible and reasonable results compared to experimental results [16,17]. It was shown that number of average photoelectrons determined at the end of whole process is reasonable if considering it with experimental ones and energy resolution values obtained are in agreement. In this study, the standard electromagnetic process was used to obtain energy deposition distribution per event. In Geant4, the processes belonging to the interactions of beams with matter are determined in seven categories: electromagnetic, hadronic, decay, photolepton-hadron, optics, parametrization, and transportation. The electromagnetic processes could be summarized as following. Photon processes include gamma conversion or pair production, photoelectric effect, Compton scattering, Rayleigh scattering, and muon pair production. Electron/positron processes cover ionization and delta-ray production, Bremsstrahlung radiation, electron-positron pair production, annihilation to two gammas of a positron, multiple scattering, the annihilation to two muons of a positron, and annihilation to two hadrons of a positron. On the other hand muon processes include the following processes: Ionization and delta-ray production, Bremsstrahlung radiation, electron/positron pair production, and multiple scattering. Hadron and ions also include ionization for hadron and ions in addition to the standard electromagnetic processes. Coulomb scattering in the model is considered different for ions and charged particles. In addition, the production of optical photons are determined with Cherenkov and scintillation processes. The standard electromagnetic package uses physics tables which are reconstructed between 100 eV and 100 TeV energy range.

HfG (Hafnium Fluoride Glass) is based on $HfF_4$-$BaF_2$-$NaF$-$AlF_3$-$YF_3$ system with molar mass fractions of 0.56, 0.28, 0.12, 0.02, and 0.02, respectively. 2.5% Ce doped HfG was used in the present study since it shows good transparency within emission spectra indicating no self-absorption. Ce doped fluorohafnate glass showed very fast decay time with short and long time constants of 8 ns and 25 ns, respectively. Its emission spectra range between 290 nm and 400 nm peaking at the wavelength of 310 nm. It has a density of 5.95 $g/cm^3$ with the refractive index of 1.495. Its radiation length is 1.6 cm and light yield of 150 photon/MeV. When it is compared to newly produced scintillation glass of Ce doped DSB [18], the following expressions could be stated: DSB glass lower stopping power. Its density and radiation length is 3.8 $g/cm^3$ and, 3.3 cm, respectively. DSB has fast decay time of 30 ns and additionally slower decay time of 180 ns. On the other hand, its light output is about five times larger than that of PWO. HfG's emission weighted transmission rate was determined as 80%. Emission weighted quantum efficiencies with APD and PIN were calculated as 43.7% and 55.5%, respectively. Transmission spectra and emission weighted quantum efficiencies together with APD and PIN diode indicates that 34.9% and 44.4% of the produced photons creates an electron in the photodetector without considering photodetector active areas.

Ce1%:$Gd_2Y_1Ga_{2.7}Al_{2.3}O_{12}$ is a new single crystal grown by Czochralski method. Its emission spectra ranges between 490 nm and 590 nm peaking at 530 nm. It reached the 65000 photons/MeV with two decay time constants of 93.5 ns and 615 ns, the relative intensities of which is 40.2% and 50.8%, respectively. Its good optical and scintillation properties together with its relatively high density of 6.3 $g/cm^3$ makes it a good alternative for gamma-ray detection and nuclear non-proliferation applications. It was seen from the report that its transmission spectra is well within its emission spectra indicating no significant self-absorption. Additionally, its emission spectra is well matched with pin diode efficiencies and APD quantum efficiency spectra. The emission weighted transmission value was determined as 79% for Ce1%:$Gd_2Y_1Ga_{2.7}Al_{2.3}O_{12}$ scintillator. Both APD and PIN diode were used as a photodetector with Ce1%:$Gd_2Y_1Ga_{2.7}Al_{2.3}O_{12}$ scintillator. Emission weighted quantum efficiencies were calculated as 84.1% and 79.5% with APD and PIN diode, respectively. When the spectra of the quantum efficiencies are considered with transmission spectra it is found that 66.5% and 62.9% of the produced photons contributes the production of photoelectron in APD and PIN diode, respectively. This will decrease when the active areas of the photodetectors are taken account.

The simulation study was performed with Geant4 high energy physics simulation package to determine the energy resolution of the interested scintillating materials as homogenous calorimeters. The intrinsic energy resolution caused by event to event energy deposition fluctuation was defined as the ratio of the sigma to the mean value of the logarithmic Gaussian fit function on the distributions of energy deposition in scintillator material per event. The fit function is given with Equation 4 [19]. Later, photodetector signal fluctuations were calculated with the appropriate process mentioned above.

$$F(x) \equiv N \exp\left(-\frac{1}{2\sigma_0^2} \ln^2\left(1 - \frac{x - x_p}{\sigma_E}\eta\right) - \frac{\sigma_0^2}{2}\right) \quad (4)$$

where $\sigma_0 = 2/\xi \sinh^{-1}(\eta\xi/2)$ and $\xi = 2\sqrt{\ln 4}$. In the formula, $x_p$ is the peak value, $\eta$ is the asymmetry parameter, $N$ is the normalization factor, and $\sigma_E$ is the full width at half maximum (FWHM) divided by $\xi$. The energy resolution was defined as the ratio of $\sigma_E$ to the peak value $x_p$.

## Results and Discussion

After this point, the name of the Ce1%:$Gd_2Y_1Ga_{2.7}Al_{2.3}O_{12}$ scintillator will be abbreviated as GdY in histograms and in the text. First of all, photoelectron production rates and the ratios of the active areas of the photodetectors to scintillator back face area were evaluated together and the average number of photoelectrons ($N_{pe}$) per MeV produced at the photodetector in an event was obtained for different scintillator backface detector geometries and photodetector combinations. This is given in Table 1.

Table 1: Number of photoelectrons per MeV ($N_{pe}$/MeV) produced for different scintillator back face areas and photodetector combinations.

| Material | Number of photoelectrons per MeV ($N_{pe}$/MeV) | | | |
|---|---|---|---|---|
| | APD | | PIN | |
| | Area (20 mm x 20 mm) | Area (25 mm x 25 mm) | Area (20 mm x 20 mm) | Area (25 mm x 25 mm) |
| GdY | 5402 | 3457 | 40881 | 26164 |
| HfG | 5,1 | 3,3 | 33 | 21 |

As expected HfG will give the lowest photoelectrons and this will cause significant contribution to energy resolution. The contributions from photodetector signal fluctuations ($a_{pe}$), which was calculated with Equation 2, are given in Table 2 for different detector combinations.

Table 2: The photostatistics parts of the parametrized energy resolution function ($a_{pe}$) with different scintillator back face areas and photodetector combinations.

| Material | Photodetector signal fluctuations ($a_{pe}$ as % in the unit of GeV$^{1/2}$) | | | |
| --- | --- | --- | --- | --- |
|  | APD | | PIN | |
|  | Area (20 mm x 20 mm) | Area (25 mm x 25 mm) | Area (20 mm x 20 mm) | Area (25 mm x 25 mm) |
| GdY | 0.066 | 0.082 | 0.016 | 0.020 |
| HfG | 1.973 | 2.467 | 0.548 | 0.685 |

As it is seen, they are very low and negligible for GdY material with both photodetectors. Those are significant for HfG scintillator. Indeed, Pin diode S1227 makes this contribution less harmful with its size and high UV sensitivity. Here, it can be said that new pin diode technology could make a scintillator more efficient compared to older photodetectors with unmatched scintillator emission spectrum at relatively lower wavelengths. A typical fit to the energy deposition distribution to obtain intrinsic energy resolution for a certain beam energy and detector geometry is shown in Figure 1. It is for the 1 GeV beam energy on the HfG calorimeter with the size of 25 mm x 25 mm back face and 27 cm in thickness. Fig. 2-5 shows intrinsic energy resolution results for all detector geometries as a function of beam energies. First of all, with evaluating all histograms the thicknesses of 17 cm and 20 cm will not be considered as material thicknesses since they do not follow a good shape with beam energies even the resolutions decrease with beam energies. For GdY and 20 mm x 20 mm cross area of each scintillator, the energy resolutions were determined as 2.14% and 1.76% for the material thicknesses of 25 cm and 27 cm, respectively at 2 GeV/c beam energy.

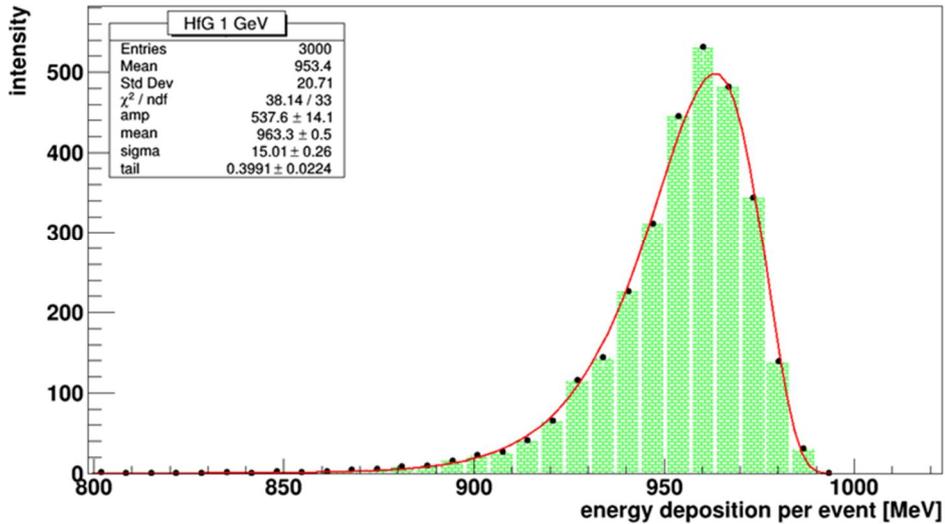

Figure 1. Typical energy resolution fitting belonging to HfG scintillator for 1 GeV beam energy.

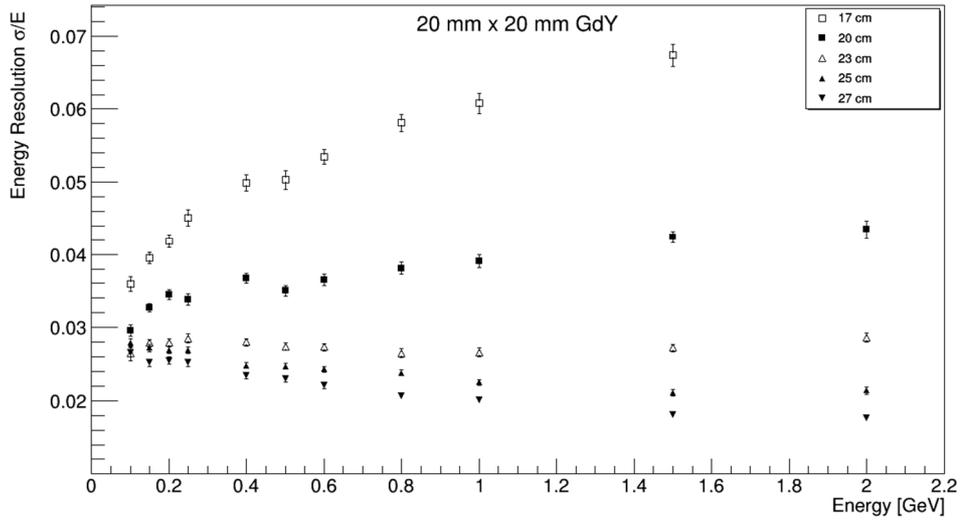

Figure 2: Energy resolutions as a function of beam energy for GdY with 20 mm x 20 mm back face area and five different thicknesses.

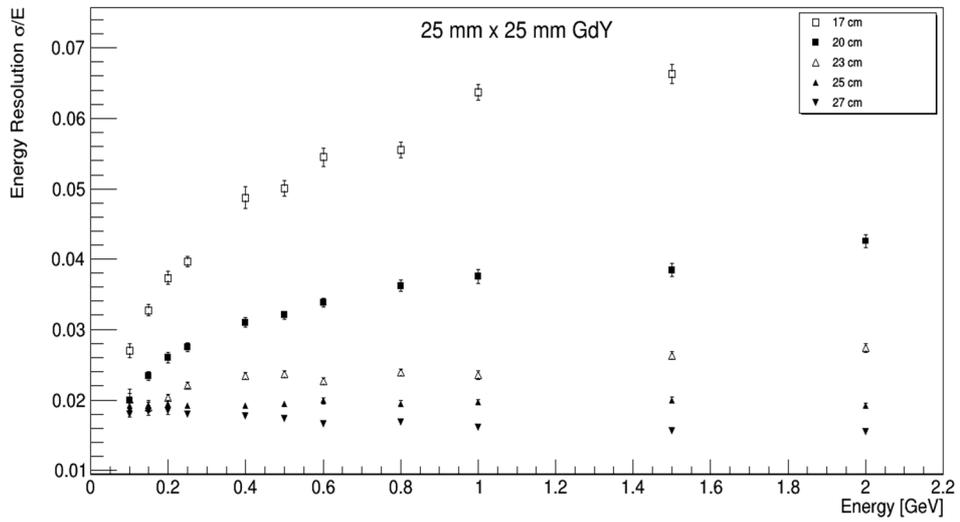

Figure 3: Energy resolutions as a function of beam energy for GdY with 25 mm x 25 mm back face area and five different thicknesses.

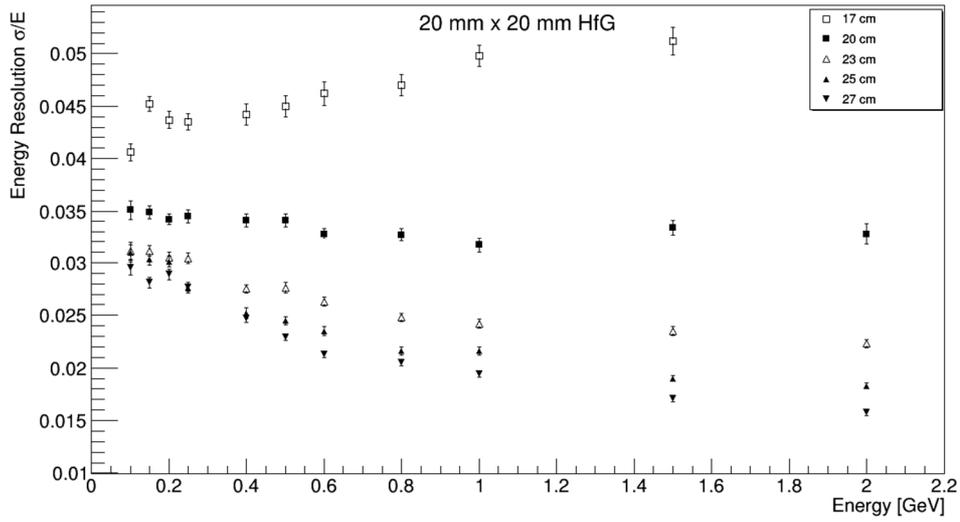

Figure 4: Energy resolutions as a function of beam energy for HfG with 20 mm x 20 mm back face area and five different thicknesses.

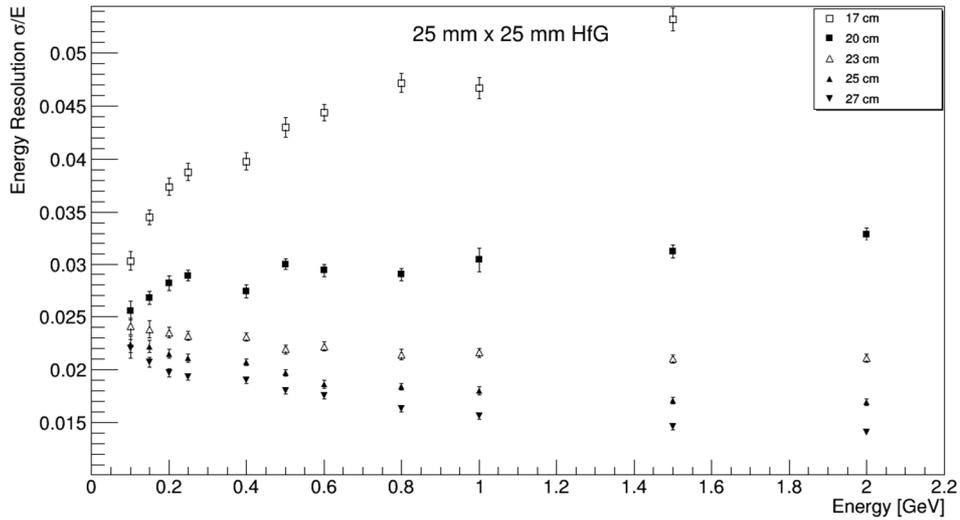

Figure 5: Energy resolutions as a function of beam energy for HfG with 25 mm x 25 mm back face area and five different thicknesses.

For GdY and 25 mm x 25 mm back face area, the resolution values has been obtained as 1.55% for 27 cm material thickness. In the case of 25 cm thick GdY, energy resolution has no proper shape fluctuating around a line with beam energy. For HfG and 20 mm x 20 mm back face area, energy resolution reached 2.23%, 1.83%, and 1.58% for 23 cm, 25 cm, and 27 cm calorimeter thicknesses, respectively. For HfG and 25 mm 25 mm cross section area of each scintillator, these values were determined as 2.11%, 1.69%, and 1.41%. It is seen that the resolutions increase with back face areas and calorimeter thicknesses. It can be stated that the resolutions belonging to the geometries of 25 mm x 25 mm beam facing area and calorimeter thicknesses of 25 cm or 27 cm gives more compatible results with previous studies [16, 17]. Therefore, energy resolution functions were parametrized with these detector geometries. Considering the selected material thicknesses and detector back face geometries, the parameters of the total energy resolution function were obtained with the function given in Equation 1 excluding noise term (c). Here, photoelectron statistics contribution was determined according to Equation 2. These parameters are given in Table 3 and Table 4 for 25 mm x 25 mm and 20 mm x 20 mm back face detector geometries, respectively. These values could not be determined for GdY material with 25 mm x 25 mm cross area and 25 cm thickness since the energy resolution values do not follow good shape with beam energies fluctuating around a line. Figures 6-8 shows parametrized energy resolution functions and resolution values calculated with the related parametrized function at certain beam energies for the scintillators with back face area of 25 mm x 25 mm and the thicknesses of 25 cm and 27 cm with APD and PIN. It should be noted that the resolutions are quite different especially at lower beam energies below 1 GeV/c whether PIN or APD is used with HfG. Finally, the followings are the best parameters obtained over the examined detector combinations with the scintillator sizes of 25 mm x 25 mm back face area and 27 cm thickness:

$$\sigma/E = 0.84\%/E^{1/4} \oplus 0.08\%/\sqrt{E} \oplus 1.39\% \quad for\ GdY + APD$$
$$\sigma/E = 0.84\%/E^{1/4} \oplus 0.02\%/\sqrt{E} \oplus 1.39\% \quad for\ GdY + PIN$$
$$\sigma/E = 1.17\%/E^{1/4} \oplus 2.47\%/\sqrt{E} \oplus 1.05\% \quad for\ HfG + APD$$
$$\sigma/E = 1.17\%/E^{1/4} \oplus 0.69\%/\sqrt{E} \oplus 1.05\% \quad for\ HfG + PIN$$

**Table 3:** Parameters of the energy resolution functions of the scintillators with different scintillator geometries and photodetector combinations.

| Material | Parameters of the energy resolution function with different scintillator thicknesses and photodetector combinations. Each scintillator has 25 mm x 25 mm back face area. | | | | | | | | | | | |
|---|---|---|---|---|---|---|---|---|---|---|---|---|
| | APD | | | | | | PIN | | | | | |
| | 25 cm | | | 27 cm | | | 25 cm | | | 27 cm | | |
| | $a_{lateral}$ | $a_{pe}$ | b | $a_{lateral}$ | $a_{pe}$ | b | $a_{lateral}$ | $a_{pe}$ | b | $a_{lateral}$ | $a_{pe}$ | b |
| GdY | - | - | - | 0.84 | 0.08 | 1.39 | - | - | - | 0.84 | 0.02 | 1.39 |
| HfG | 1.09 | 2.47 | 1.44 | 1.17 | 2.47 | 1.05 | 1.09 | 0.69 | 1.44 | 1.17 | 0.69 | 1.05 |

**Table 4:** Parameters of the energy resolution functions of the scintillators with different scintillator geometries and photodetector combinations.

| Material | Parameters of the energy resolution function with different scintillator thicknesses and photodetector combinations. Each scintillator has 20 mm x 20 mm back face area. | | | | | | | | | | | |
|---|---|---|---|---|---|---|---|---|---|---|---|---|
| | APD | | | | | | PIN | | | | | |
| | 25 cm | | | 27 cm | | | 25 cm | | | 27 cm | | |
| | $a_{lateral}$ | $a_{pe}$ | b | $a_{lateral}$ | $a_{pe}$ | b | $a_{lateral}$ | $a_{pe}$ | b | $a_{lateral}$ | $a_{pe}$ | b |
| GdY | 1.25 | 0.07 | 1.91 | 1.41 | 0.07 | 1.40 | 1.25 | 0.02 | 1.91 | 1.41 | 0.02 | 1.40 |
| HfG | 1.80 | 1.97 | 1.07 | 1.83 | 1.98 | 0.57 | 1.80 | 0.55 | 1.07 | 1.83 | 0.55 | 0.57 |

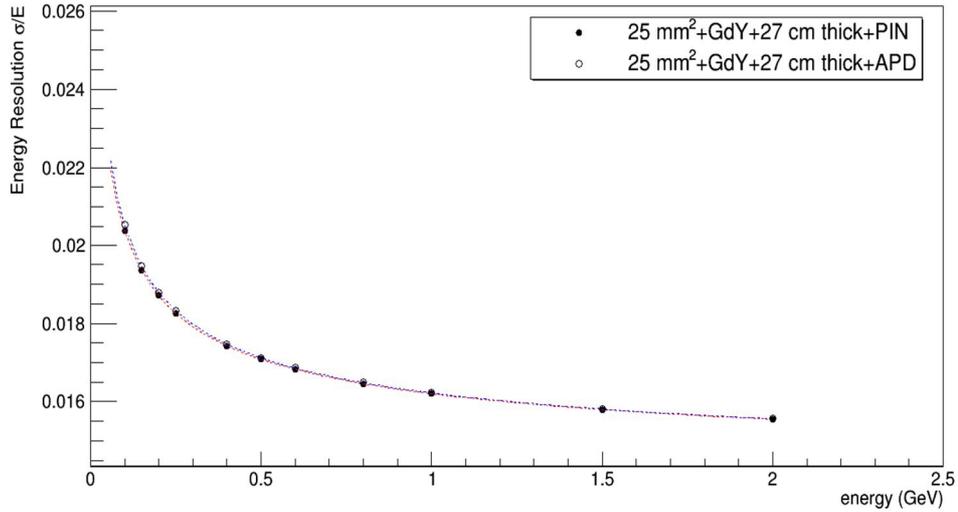

Figure 6: Parametrized energy resolution function (dashed lines) and energy resolution values at certain energies (points) for scintillator–photodetector combinations belonging to GdY with 25 mm x 25 mm back face area of each scintillator and 27 cm thickness.

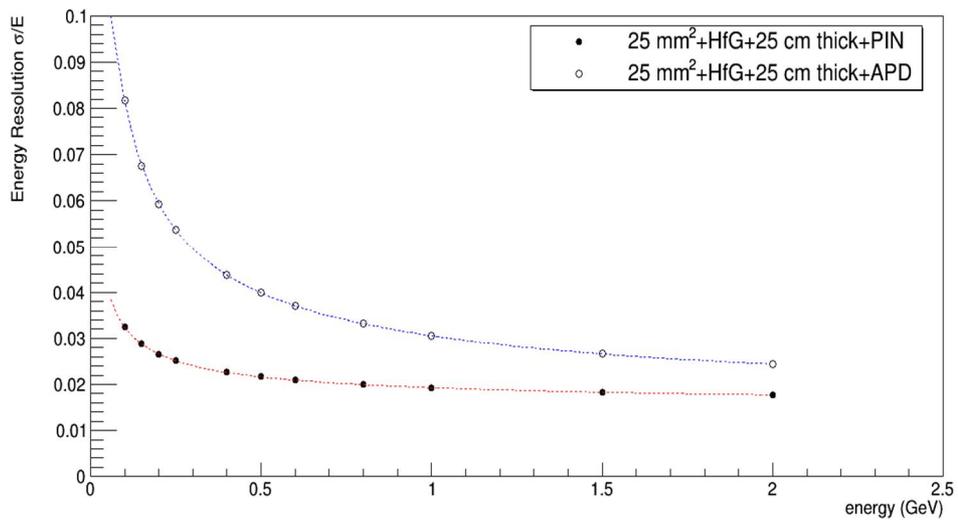

Figure 7: Parametrized energy resolution function (dashed lines) and energy resolution values at certain energies (points) for scintillator–photodetector combinations belonging to HfG with 25 mm x 25 mm back face area and 25 cm thickness.

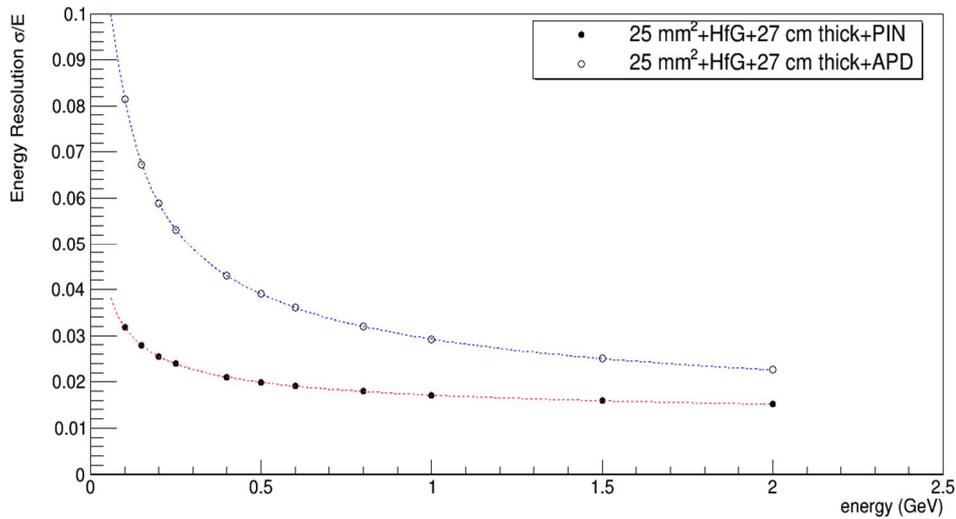

Figure 8: Parametrized energy resolution function (dashed lines) and energy resolution values at certain energies (points) for scintillator–photodetector combinations belonging to HfG with 25 mm x 25 mm back face area and 27 cm thickness.

After this point, the same parametrized results were obtained as 3x3 matrix with the optimized detector geometries of 27 cm in thickness and 25 mm x 25 mm back face area for scintillator-PIN diode combinations. The results were shown together with 5x5 matrices for GdY and HfG in Figure 9 and 10, respectively. It is obvious that energy resolutions increase with transverse sizes. It is also seen that transverse size is more effective at lower beam energies below 1 GeV/c.

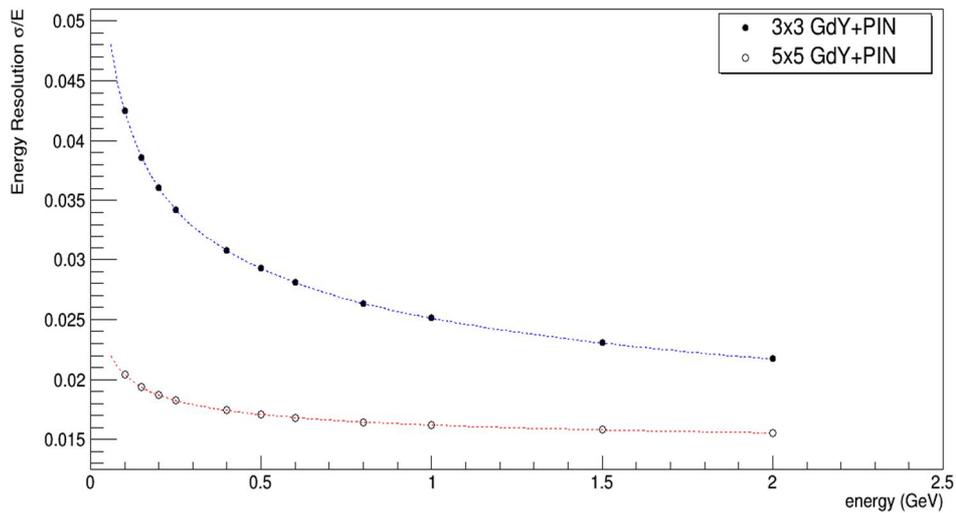

Figure 9: Parametrized energy resolution function for GdY as 3x3 and 5x5 matrices. Each scintillator has 27 cm thickness and back face area of 25 mm x 25 mm.

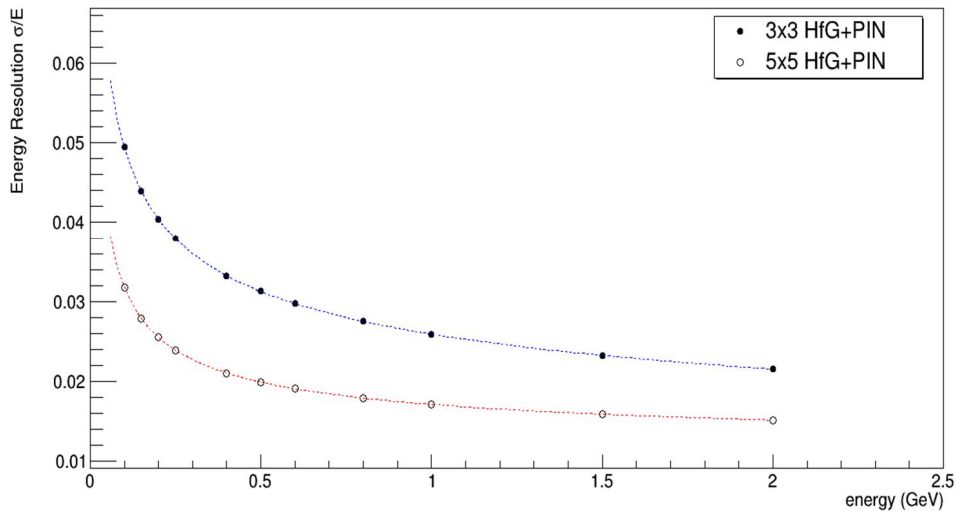

Figure 10: Parametrized energy resolution function for HfG as 3x3 and 5x5 matrices. Each scintillator has 27 cm thickness and back face area of 25 mm x 25 mm.

## Conclusions

A computational study was carried out to determine energy resolutions of two different scintillator materials to be used as a homogenous calorimeter in particle physics experiments. Since GdY has very high light yield and its emission spectrum matches well with two photodetectors, photostatistics contribution to the total energy resolution is negligible. For HfG, in both case there will be significant contribution but with APD this will be enormous resulting huge decrease on energy resolution especially at lower beam energies below 1 GeV/c. Above 1 GeV/c, this contribution could be thought as reasonable. On the other hand, it could be stated that the resolution will increase significantly at lower beam energies below 1 GeV/c if the number of photodiodes are increased. In addition, the calculation procedure of the average number of photoelectrons will give the estimation of the minimum number of photons detected. In a real experiment, the possibility of detection of the number of the photons will increase due to randomly polarized photons and scattering via surface reflectors. If PIN diode is used with HfG, this contribution will be very reasonable especially at relatively high energies. It could be stated that both scintillators will give very compatible results for material thicknesses of 25 or 27 cm and with appropriate photodetectors when compared to previous studies. This allows to be stated that these materials could be seen as alternatives in particle physics experiments by taking account their advantages. GdY's very high light yield could make it preferable especially for relatively low beam energies. The main advantage of HfG is its very fast decay times of 8 ns and 25 ns. Finally, their relatively high densities allows to reach compatible resolutions results with smaller detector sizes compared to scintillators with lower densities.

## Data Availability

The simulation data used to support the findings of this study are available from the corresponding author upon request.

## Conflicts of Interest

The author declares that there is no conflict of interest regarding the publication of this paper.

## Funding Statement

The author declares that there is no funding body for this study.

## Acknowledgments

The author would like to thank to Dr. Ugur Akgun for his valuable comments on this work.

## Supplementary Materials

The author declares that there is no supplementary materials for this paper.

## References

[1] R. Paramatti on behalf of the CMS Collaboration, Design options for the upgrade of the CMS electromagnetic calorimeter, Nuclear and Particle Physics Proceedings 273–275 (2016) 995–1001.

[2] CMS Collaboration, "The CMS experiment at the CERN LHC", *JINST* 3 (2008) S08004.

[3] E. Auffray, et al., Cerium doped heavy metal fluoride glasses, a possible alternative for electromagnetic calorimetry, Nuclear Instruments and Methods in Physics Research Section A 380 (1996) 524-536.

[4] I.Dafinei, E. Auffray, P.Lecoq, M.Schneegans: MRS Proceedings Series (Scintillators and Phosphors), vol. 348 (1994) p.217.

[5] Chao Wang, et al., Optical and scintillation properties of Ce-doped $(Gd_2Y_1)Ga_{2.7}Al_{2.3}O_{12}$ single crystal grown by Czochralski method, Nuclear Instruments and Methods in Physics Research Section A 820 (2016) 8-13.

[6] R.Y. Zhu, Crystal calorimeters in the next decade, Journal of Physics: Conference Series 160 (2009) 012017.

[7] B. Lewandowski, The BaBar electromagnetic calorimeter, Nuclear Instruments and Methods in Physics Research Section A 494 (2002) 303.

[8] K. Miyabayashi, Belle electromagnetic calorimeter, Nuclear Instruments and Methods in Physics Research Section A 494 (2002) 298.

[9] M. Ablikim, Z. H. An, J. Z. Bai, et al., Design and construction of the BesIII detector, Nuclear Instruments and Methods in Physics Research Section A 614 (2010) 345.

[10] D. Renker, Properties of avalanche photodiodes for applications in high energy physics, astrophysics and medical imaging, Nuclear Instruments and Methods in Physics Research Section A 486 (2002) 164.

[11] URL:http://www.hamamatsu.com

[12] E. Pilicer, F. Kocak, I. Tapan, Excess noise factor of neutron-irradiated silicon avalanche photodiodes, Nuclear Instruments and Methods in Physics Research Section A 552 (2005) 146.

**[13]** S. Agostinelli et al., Geant4—a simulation toolkit, Nuclear Instruments and Methods in Physics Research Section A 506 (2003) 250-303.

**[14]** J. Allison et al., Geant4 developments and applications, IEEE Transactions on Nuclear Science 53 No. 1 (2006) 270-278.

**[15]** J. Allison et al., Recent developments in Geant4, Nuclear Instruments and Methods in Physics Research Section A 835 (2016) 186-225.

**[16]** F. Kocak, Simulation studies of crystal-photodetector assemblies for the Turkish accelerator center particle factory electromagnetic calorimeter, Nuclear Instruments and Methods in Physics Research A 787 (2015) 144–147.

**[17]** F. Kocak and I. Tapan, Simulation of LYSO Crystal for the TAC-PF Electromagnetic Calorimeter, ACTA PHYSICA POLONICA A, Vol. 131 (2017) 527.

**[18]** E Auffray et al., DSB:Ce3+ scintillation glass for future, J. Phys.: Conf. Ser. 587 (2015) 012062.

**[19]** H. Ikeda et al., A detailed test of the CsI(Tl) calorimeter for BELLE with photon beams of energy between 20 MeV and 5.4 GeV, Nuclear Instruments and Methods in Physics Research A 441 (2000) 401-426.